# Privacy as Contextual Integrity in Online Proctoring Systems in Higher Education: A scoping review

Chantal Mutimukwe
Stockholm university, Sweden
chantal.mutimukwe@dsv.su.se

Shengnan Han
Stockholm University, Sweden
shengnan@dsv.su.se

Olga Viberg
Royal Institute of Technology (KTH), Sweden
oviberg@kth.se

Teresa Cerratto-Pargman
Stockholm University, Sweden
tessy@dsv.sus.e

## Abstract

*Privacy is one of the key challenges to the adoption and implementation of online proctoring systems (OPS) in higher education. To better understand this challenge, we adopt privacy as contextual integrity theory to conduct a scoping review of 17 papers. The results show different types of students' personal and sensitive information are collected and disseminated; this raises considerable privacy concerns. As well as the governing principles including transparency and fairness, consent and choice, information minimization, accountability, and information security and accuracy have been identified to address privacy problems. This study notifies a need to clarify how these principles should be implemented and sustained, and what privacy concerns and actors they relate to. Further, it calls for the need to clarify the responsibility of key actors in enacting and sustaining responsible adoption and use of OPS in higher education.*

## 1. Introduction

The use of online proctoring systems (OPS) in higher education has been rapidly evolving during the last few years (Han et al., 2022). The motivation for adopting these tools lies in their perceived ability to provide integrity, authentication, authorization, and operational control of online exams in remote environments (Nigam et al., 2021). Scholars posit that online proctoring technology will become the "new normal" in higher education (Selwyn et al., 2021). However, a few universities, such as Oxford, Michigan-Dearborn and Cambridge rejected to use OPS because of different challenges related to OPS, where student privacy is paramount (Coghlan et al., 2021; Silverman et al., 2021).

Recent studies emphasize that student privacy is one of the main challenges to the adoption and implementation of OPS in higher education (Baume, 2019; González-González et al., 2020; Nigam et al., 2021). Although, privacy issues in relation to the use of OPS in higher education have not been comprehensively studied (Balash et al., 2021; Coghlan et al., 2021). The existing studies on OPS focused on the challenges related to students' perceptions of OPS, students' performance, anxiety, cheating problems, staff perceptions, authentication and exam security, interface design, as well as technology-related issues; yet they have barely discussed privacy issues in particular (Butler-Henderson & Crawford, 2020; Nigam et al., 2021). It is worthy to note that addressing privacy problems in a certain context, such as in the setting of higher education is crucial to comprehending the aspects of privacy in that context (Wu, 2014). To this end, we choose the lens of privacy as contextual integrity theory (Nissenbaum, 2004; 2010) to provide an understanding of privacy issues in relation to the use of OPS in the context of higher education.

We conducted a scoping review of the peer-reviewed studies focusing on privacy aspects when using OPS in higher education. Through the lens of privacy as contextual integrity theory, this study has analyzed the 17 identified articles to answer the following research questions:
1. What are the information types that are collected through the use of OPS in higher education?
2. What are the roles of actors involved in the process of information flow in OPS in higher education?
3. What are the principles to govern the information flows in OPS in higher education?

The paper is structured as follows: section 2 presents a literature background on the OPS and privacy; section 3 presents privacy as contextual integrity; section 4 introduces the research method; the results are reported in section 5. Finally, the discussion is presented in section 6, and the conclusion is drawn at the end.

## 2. Background

### 2.1 Online Proctoring Systems (OPS) in higher education

'Online proctoring' also known as 'digital 'proctoring' or 'remote proctoring' is defined as "the process of using digital tools and technologies to ensure that candidates taking examinations and other forms of assessment comply with prescribed policies and guidelines" (Udechukwu, 2020, p.6262). Consequently, OPS promise to allow students and course participants to take their exams anywhere in a secure and reliable way (Baume, 2019).

Many OPS have been developed, including ConductExam, Honorlock, IRIS, Mercer Mettl, ProctorExam, Proctorio, ProctorU and PSI Online (Arnò et al., 2021). OPS can be divided into three categories (Arnò et al., 2021; Nigam et al., 2021): 1. live proctoring, 2. recorded proctoring, and 3. automated proctoring. Live proctoring involves real-time proctoring taking place during the exam with a human proctor monitoring/supervising the exam virtually or online (Nigam et al., 2021). Recorded proctoring does not utilize a human invigilator, but instead, student behavior is recorded during the examinations (Nigam et al., 2021). Automated proctoring consists of OPS human proctors, and does not monitor the entire exam; instead, the proctoring system identifies key events of possible fraud or cheating (Nigam et al., 2021).

OPS have received attention in academic research across various disciplines. Nigam et al. (2021) for example, reviewed artificial intelligence (AI)-related features that are largely used in various digital proctoring systems. They then addressed four primary research questions focusing on the existing architecture of proctoring systems, parameters to be considered for OPS, trends, issues in OPS, and the future of OPS. Based on a review of 29 OPS, Arnò et al. (2021) summarized the state-of-the-art proctoring systems by identifying and describing their main features and analyzing the way in which different proctoring programs are grouped on the basis of the services they offer. Further, Butler-Henderson and Crawford (2020) conducted another systematic review that explored the challenges and opportunities of online examination, including OPS. They analyzed 56 articles and explored the following key themes: student perceptions, student performance, anxiety, cheating, staff perceptions, authentication and security, interface design, and technology issues. While the literature on OPS is growing, and some studies pointed to privacy as one of the critical challenges of the adoption and implementation of OPS (González-González et al., 2020; Nigam et al., 2021), there is still a lack of in depth-discussion of privacy issues related to the use of OPS in higher education settings.

### 2.2 Privacy

Privacy is understood differently in research and practice. Smith et al. (2011) classify privacy into two categories: value-based and cognate-based. The value-oriented class considered privacy as a 'right' or 'commodity'. The concept of privacy as a 'right' was introduced by Warren and Brandies (1890) in their seminal essay, in which they defined privacy as the right of the individual to be left alone and free from intrusion and interference. The notion of privacy as a commodity suggests that privacy remains an individual and social value, not an absolute right, but subject to the economic principles of cost-benefit analysis and trade-offs (Bennett, 1995).

The cognate-based definitions explain privacy in two ways: privacy as a 'state' and privacy as 'control'. The concept of privacy as a 'state' was introduced by Westin (1967), who defined privacy as four different sub-states: anonymity, solitude, seclusion, and intimacy. Later, Schoeman (1984) also defined general privacy as "a state of limited access to a person". The concept of privacy as 'control' is based on Westin's (1967) and Altman's (1975) theories of privacy. Margulis (1977) further elaborated on Westin and Altman's perspectives and proposed a control-centered privacy definition: "Privacy, as a whole or in part, represents the control of transactions between person(s) and other(s), the ultimate aim of which is to enhance autonomy and/or to minimize vulnerability" (p.10).

The information systems research field largely relies on the theory of privacy as control or state (Mutimukwe et al., 2020, 2022). However, these theories lack explanatory power when it comes to shedding light on the boundaries of a specific context, specially drawn between public and private information in actual online practices in a digital age (Bélanger & Crossler, 2011). The world of information technology is heterogeneous; the differences between public and private information are often shaped by the nature of the context, which implies that a single privacy theory or framework could not be applicable to all contexts (Nisenbaum, 2004).

## 3. Privacy as contextual integrity theory

Nissenbaum (2004, 2010) proposed the theory of *privacy as contextual integrity* to bridge the gaps in the earlier theories. It can be considered as an alternative benchmark for privacy to capture the privacy issues posed by information technology tools (Nissenbaum, 2004), such as OPS. It is one of the most influential theories explaining the privacy issues that can be observed along with the development of a new technological tool (Hoel et al., 2020) such as an OPS. Nisssenbaum (2010) argues that:

"a right to privacy is neither a right to secrecy nor a right to control but a right to appropriate flow of personal information [...] Privacy may still be posited as an important human right or value worth protecting through law and other means, but what this amount to is contextual integrity and what this amount to varies from context to context" (p. 127).

In sum, the theory of privacy as contextual integrity suggests that privacy is about the appropriate flow of information in a certain context. "It ties adequate protection for privacy to norms of specific contexts, demanding that information gathering and dissemination be appropriate to that context and obey the governing norms of distribution within it" (Nissenbaum, 2004, p.119).

The norms of a specific context can be depicted by five key parameters: 1. **Information subject,** 2. **Sender** and 3. **Receiver**, 4. **Attributes**, and 5. **Transmission principles,** which refer to *"the constraints on the flow of information from party to party in a context"* and the "terms and conditions under which such transfers should occur" (Nissenbaum, 2010, p.132). Here, the *Sender*, the *Recipient,* and the *Information subject* refer to actors' roles and attribute to the type of information exchange under a specified transmission principle. For example, in the U.S. education context, where information exchange between actors is regulated by the Family Educational Rights and Privacy (FERPA) policies (US Department of Education, 2021), the teacher (the sender role) would be allowed to share grades (type of information) of a student (the information subject role) with her parents (the receiver role), if the student has given explicit written permission (transmission principle).

The description of the five key norms parameters is a paramount way to illustrate the problems that can occur in a certain context and hence the problems in the breadth and depth of privacy issues in that context (Heath, 2014). Hence, this study provides an increased understanding of different privacy issues related to the use of OPS in higher education, with the focus of describing: i) the types of information that are collected in OPS and related privacy concerns, ii) different actors that could be involved in the process of information collection and dissemination, and iii) transmission principles that can guide the information flow in OPS in higher education context.

Nissenbaum defines *contexts* as "structured social settings characterized by canonical activities, roles, relationships, power structures, norms (or rules), and internal values (goals, ends, purposes)" (2010, p. 132). *OPS in higher education* is a context in which the acknowledged activities, roles, and relationships are evident. The student takes an online exam, and educators and academic staff may: monitor the exam online, analyze the recorded student's behavior, and collect evidence of possible fraud or cheating.

The internal values –goals, ends, and purposes – in the OPS and higher education context relate to the provision of integrity, authentication, authorization, and operational control of online exams. As students engage with OPS, data is generated as a by-product of this activity and does provide valuable insights into student engagement in online exams.

## 4. Research method

We conducted a scoping literature review, which is of "a particular use when the topic has not been extensively reviewed or is of complex or heterogeneous nature" (Pham et al., 2014, p.371). Scoping literature review is correspondingly suitable if the study aims at clarifying a specific concept within the literature (Munn et al., 2018).

The literature review considered all related peer-reviewed articles across various disciplines. We retrieved data from five databases: *ACM, ERIC, IEEE, Scopus* and *Web of Science*, and considered articles that were published during the last five years, between 2018-2022. The retrieval was based on search keywords and the following boolean operators: ('privacy' OR 'security' OR 'ethics') AND ('proctoring' OR 'invigilation' OR "online exam" OR "remote exam" OR "digital exam" OR "online assessment" OR "digital assessment" OR "remote assessment") AND ("higher education" OR university OR college). Although we focused on privacy specifically, we also considered such keywords as 'ethics' and 'security' as some studies consider privacy as one of the securities or ethical dimensions (Smith et al., 2011).

The literature search process was based on the PRISMA process (Page et al., 2021; Figure 1). It was conducted and completed between April 11th and May 16th, 2022, and we initially found 197 articles in total (Figure 1). After removing the duplicates (n=55), and the papers that are not written in English (n=1), 141 papers remained. We followed the following inclusion/exclusion criteria to screen the relevance of the articles. (1) The article must study online proctoring systems. (2) It should be conducted within higher education settings. 3) It discusses privacy in an explicit or implicit way. (4) In the study, the proctoring system is used for online examination or assessment purposes. We have manually checked all the 141 articles in the whole text, and finally, 17 articles were considered to meet the criteria as relevant to the literature review and included in the final analysis.

Among the 17 papers, 11 articles were published in 2021. Two articles were published in 2020 and 2022, respectively. For 2019 and 2018, we found one article published each year. In terms of research methodology, five articles were conceptual (n=5). Quantitative (n=1), qualitative (n=2), mixed methods (n=3), mathematical modeling (n=3), and design (n=2) were also used in the analyzed articles. One literature review also appeared. Eleven articles indicated a country, in which the study was conducted and/or from which the data were obtained. Six studies were conducted in European countries; other studies were performed in other countries, including Russia (n=1), the United States (n=2), Australia (n=1), and the United Arab Emirates (n=1).

**Figure 1**

*PRISMA Diagram of the literature search*

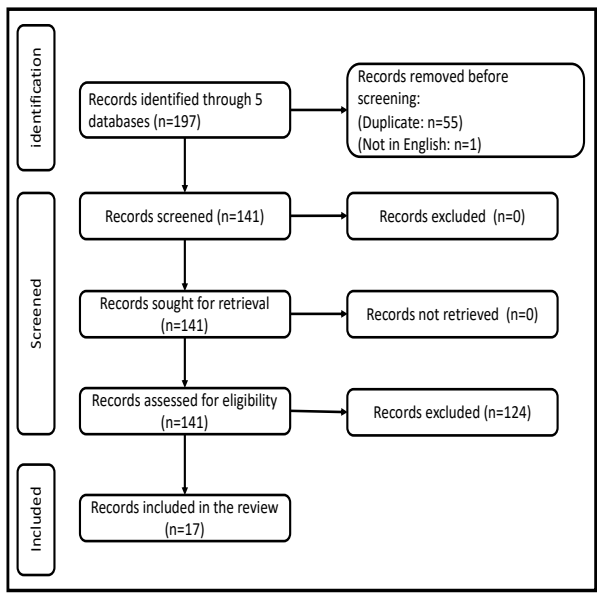

Only four articles (Balash et al., 2021; Coghlan et al., 2021; Henry & Oliver, 2022; Kharbat & Daabes, 2021) thoroughly and explicitly discussed privacy issues. Balash et al. (2021) and Kharbat and Daabes (2021) examined students' perceptions of OPS and discussed in detail students' privacy concerns regarding OPS; they suggested some recommendations and guidelines address these concerns. Coghlan et al. (2021) as well as Henry and Oliver (2022) addressed the ethical aspects of OPS and discussed privacy at length, among other ethical aspects. Others (n=13) implicitly mentioned or discussed privacy among other challenges of OPS's implementation (Baume, 2019; Draaijer et al., 2018; González-González et al., 2020; Labayen et al., 2021), acceptance (Duric & Mahmutovic, 2021; Elshafey et al., 2021; Langenfeld, 2020; Meulmeester et al., 2021; Purohit & Ajmera, 2022; Silverman et al., 2021; Tripathi et al., 2022), trust (Nigam et al., 2021), and OPS's negative aspects (Bundin et al., 2021).

For the analysis, we used the thematic analysis method (Braun & Clarke, 2006) to interpret the key insights from the selected articles. Embracing a theory-driven approach and using privacy as contextual integrity theory as the theoretical lens, we followed the six-step model described by Braun and Clarke (2006) in the data analysis process: 1) Familiarisation with data. The papers included in the final review were read more than once to become familiar with them and to have a clear understanding. 2) Generating initial codes. The initial codes were generated based on the definition/understanding of privacy. This was an iterative process that lasted until no new codes were identified. 3) Searching for themes among codes. As we applied a theory-driven approach, the themes identified as "the type of information collected", "roles of the key actors" and "governing principles". 4) Reviewing themes. The themes were reviewed to ensure that they were supported by the coded data. 5) Defining and naming themes. We refined and described our findings. 6) Producing the final report. To ensure the reliability and validity of the results, the team performed the steps jointly. We achieved consensus and agreements in interpretations through continuous discussions.

## 5. Results

### 5.1 Type of information collected in OPS in higher education

We found that the use of OPS in higher education can enable the collection of a wide range of sensitive and personal information about students. We classified this information into three categories: ***1. Individual identification information, 2. Monitoring and controlling devices' information, and 3. AI-based and biometric information.***

Information related to the individual's identification mostly includes students' full names, personal identification numbers, email addresses, phone numbers, student ID numbers, education institutions, e-mail addresses, driving license, passport numbers, home address, and birth date (e.g., Balash et al., 2022; Bundin et al., 2021; Coghlan et al., 2021; Draaijer et al., 2018; Duric & Mahmutovic, 2021).

Monitoring and controlling the devices' information include screen images, accessing web page content, blocking browser tabs, browser history, analysis of keyboard strokes (keystrokes or syntax), and changing privacy settings, IP address (e.g., Draaijer et al., 2018; Henry & Oliver, 2022; Kharbat & Daabes, 2021; Labayen et al., 2021; Langefled, 2020; Purohit & Ajmera, 2022; Silverman et al., 2021).

AI-based and biometrics information include facial and voice detection, tracks of eyes movement, track of student's behavior, room and home scan with a 360-degree camera; caption of the upper body, hands, and desk (using profiled angle camera) (e.g., Balash et al., 2021; Baume, 2018; Bundin et al., 2021; Coghlan et al., 2021; Draaijer et al., 2018).

## 5.2 The roles of different actors in OPS in higher education

By examining the 17 articles, we identified six actors in the context of using OPS in higher education, which include **(1)** *students* (e.g., Balash et al., 2021; Baume, 2018; Bundin et al., 2021; Coghlan et al., 2021; Purohit & Ajmera, 2022; Silverman et al., 2021), **(2)** *teachers or educators* (e.g., Coghlan et al., 2021; González-González et al., 2020; Henry & Oliver, 2022; Kharbat & Daabes, 2021; Labayen et al., 2021; Purohit & Ajmera, 2022; Silverman et al., 2021) **(3)** *institutions* (e.g., Balash et al., 2021; Baume, 2018; Bundin et al., 2021; Coghlan et al., 2021; Henry & Oliver, 2022; Kharbat & Daabes, 2021), **(4)** *commercial technology companies* (e.g., Balash et al., 2021; Bundin et al., 2021; Coghlan et al., 2021; Henry & Oliver, 2022), **(5)** *regulators and responsible authorities at the national level* (Bundin et al., 2021), and **(6)** *AI-human actors* (Henry & Oliver, 2022). The major concern is that the roles of the key actors have not yet been discussed in the reviewed sample. It is clear that students are the main information subjects. 'Information subject' is defined as identified or identifiable nature person (EU-GDPR, 2018a). In other words, an information subject is a person whose personal information is collected, held and processed by another entity. Personal information is any information that can be used to identify an individual such as name, ID, etc. (EU-GDPR, 2018a). The students are 'information subjects' as all other actors have interest in their information. Yet, the roles of other actors are still ambiguous, and it is not clear yet who is the main *sender* or the *receiver* of information.

## 5.3 The principles to govern information flow in OPS in higher education

'Transmission principles' is one of the key five parameters of contextual norms. They are defined as "the constraints on the flow of information from party to party in a context" and the "terms and conditions under which such transfers should occur" (Nisenbaum, 2010, p. 132). By thoroughly examining the selected sample, we found five principles that were frequently mentioned or identified to be appropriate for governing the information flow in OPS context: **1)** *Transparency and fairness*, **2)** *information minimization,* **3)** *Consent and choice,* **4)** *Information security and accuracy, and* **5)** *Accountability.*

**Transparency and Fairness:** Transparency generally implies that the different actors, especially the information subject, should be informed on how their data is being used (EU-GDPR, 2018b). It is inherently linked to fairness, and this involves being clear, open, and honest with information subjects about who is accessing their data, and why and how their data are being processed (EU-GDPR, 2018b).

Some of the examined articles discussed or mentioned that transparency and fairness are important to make different actors comfortable, reduce the students' privacy concerns, and increase the acceptance and the use of OPS (Balash et al., 2021; Kharbat & Daabes, 2021). Balash et al. (2021) noted that students should be informed about the privacy implication of using OPS, and the clear rationale for using OPS. They should also get the notice before online proctored exams, and clear instructions and/or assistance in removing invasive monitoring software following an exam. Syllabi could include a readable privacy policy to better communicate expectations (Balash et al., 2021). Coghlan et al. (2021) also pointed out that OPS-technology companies and universities should be transparent about how the OPS technology works, how it will be used in particular circumstances, and what impact they may have on the students. For example, students should be informed that they are not being leered at or that online proctors have not shared their images with third parties (Coghlan et al., 2021). Gonzalez-Gonzalez et al. (2020) also recommended that technology companies must carry out good communication and awareness campaigns regarding the privacy of the OPS if they want to conquer and consolidate its use in online teaching institutions. Meulmeester et al. (2021) stressed that privacy issues may be solved by improved communication about OPS functionalities and the involvement of human judgment in the interpretation of the data generated by the OPS.

**Accountability:** The accountability principle implies that different actors must be accountable for the proper processing of personal information and compliance with the terms of conditions under the context (EU-GDPR, 2018b). In the context of OPS in higher education, accountability implies that all actors should be consulted and adequately informed about the impacts and capabilities of selected OPS. For example, Balash et al. (2021) posit that the students, along with faculty and administrators, take part in the assessment and selection of exam proctoring software. Students should be involved in every step of the process, from deciding whether to use exam proctoring software to determining which, if any, software should be used and which methods should be made available for exam monitoring. Bundin et al. (2021) correspondingly highlight that the participation of national regulators and responsible authorities will be required especially for evaluating existing OPS biometrics solutions on the market and recommendations for educational institutions.

**Consent and choice:** The principle of consent and choice generally considers presenting the information subject with the choice or not to allow the processing of his or her personal information (ISO, 2015). Concerning the OPS in higher education, Balash et al. (2021) indicated that providing some form of consent to students before an online proctored exam can ease their privacy concerns. Correspondingly, Coghlan et al. (2021) noted that the collection of biometric information without students' consent is clearly unethical. They also argued that:

> "A robust standard of genuine consent would also allow students to be able not to consent to OPS without penalty and to freely choose instead a human invigilator. If this option is unavailable, then the consent cannot be considered genuine consent" (p.1597).

Balash et al. (2021) also pointed out that offering students more choices for assessment and being upfront with students about institutional privacy norms is a crucial step to alleviating privacy concerns.

**Information minimization:** this considers the minimization of personal information which is processed (ISO, 2015). For OPS used in a higher education context, limiting the use of biometrics information, where it is undoubtedly necessary and inevitable, will ease the challenges of OPS including privacy issues (Bundin et al., 2021). Balash et al. (2021) moreover recommended that universities and educators would adopt the approach of least monitoring by using the minimum number of monitoring types necessary, by considering the class size and expected student behavior.

**Information security and accuracy:** Information security and accuracy require ensuring that personal information is secured and accurate. To reduce security and improper access problems in OPS, students' captured information can be stored in encrypted form on third parties host servers (Coghlan et al., 2021).

# 6. Discussion

Our findings show that the use of OPS in the higher education context is associated with the collection and dissemination of a wide range of sensitive and personal student information, including the information related to individual identification, device monitoring and control, and AI-based and biometric information. The collection of such types of information is privacy invasive, and some of these types of information may in principle not be processed unless the law provides specific or general exceptions in some the European countries (Draaijer et al., 2018). This raises significant privacy concerns (Balash et al., 2021; Kharbat & Daabes, 2021) which could lead to anxiety, distrust in these systems, student performance issues, or test taker withdrawal (e.g., Balash et al., 2021; Langenfeld, 2020). Smith et al. (1996) revealed four relevant categories of individual concerns about institution privacy practices; 1) Information collection, 2) unauthorized secondary use of personal information, 3) improper access, and 4) errors. The concerns that are related to the collection of information in OPS in higher education fall into all of these categories.

**'Information collection'** involves an individual's concern about the amount of personal information that is collected and stored (Smith et al., 1996). For the use of OPS, students express concern about the amount of personal information collected (Balash et al., 2021; Kharbat & Daabes 2021; Henry & Oliver, 2022). For example, the use of 360-degree camera to scan the students' home or rooms, and voice detection tools cause students' concerns about disclosing too much and unnecessary information about their private lives, including visual and audio materials about their private residences.

**'Unauthorized secondary use'** takes into account the individual's concern that personal information collected may be used for another purpose without the information subject's consent (Smith et al., 1996). Students have expressed concerns that in some of the OPS, it is possible for third parties to transmit the video and audio information without the students' awareness or consent (Coghlan et al., 2021). It is important to note that the information in OPS is frequently managed by third-party entities, and students do not know what could happen to their information or how long it is stored (Balash et al., 2021; Coghlan et al., 2021). Another important concern is that scanning and sharing important documents such as ID cards, passports, and driver's licenses often involve sensitive student information and which can be easily misused (Coghlan et al., 2021). In addition, sharing cell phone numbers can also lead to phishing calls and serious crimes such as catfishing, harassment, etc. (Coghlan et al., 2021).

**'Improper access'** refers to an individual's concern that personal information is readily available to people who are not properly authorized to view or work with that information (Smith et al., 1996). Some OPS scan and monitor students' computers and this can lead to improper access of sensitive and personal information (Balash et al., 2021, p.642). Coghlan et al.'s study (2021) also emphasizes that monitoring students' computers can lead to personal or sensitive information being inadvertently captured, pointing to a case where a student's credit card information was inadvertently displayed on the screen. Another issue is that students sometimes need to share their webcams in order to scan their rooms and apartments, which can lead to personal information about family members being captured without their consent (Balash et al., 2021; Henry & Oliver, 2022; Kharbat & Daabes, 2021).

**'Errors'** is about individuals' concern that there are insufficient protections against intentional and accidental errors in handling personal information (Smith et al., 1996). Students also worry about the accuracy and protection of personal information with the host, which is mainly external companies (Langenfeld, 2020).

On this basis, we can argue this study justifies the rejection of OPS in various universities (Coghlan et al., 2021; Silverman et al., 2021;), and critics claim that the use of OPS is technology-centric rather than human-centric (Henry & Oliver, 2022). Yet, previous studies suggest that instead of simply rejecting OPS due to different challenges including privacy concerns, researchers and practitioners may closely cooperate to develop new insights, conditions, and principles into how these tools might enact (Henry & Oliver, 2022). Our findings point out five principles: 1. transparency and fairness, 2. consent and choice, 3. data minimization, 4. accountability, as well as 5. information security and accuracy to be most recommended or mentioned to serve as terms of conditions of information flow in OPS context. Although these principles have been identified to be important, especially transparency, fairness, and accountability, some scholars criticize them to be ethical washing principles and only appease technology companies' interests (Henry & Oliver, 2022). Larsson (2020) posited for example, that they are mostly interpreted in a vague way, and it is not clear why they are considered important, what concerns to address, and what actors they relate to. He further suggested these principles need to consider the issues of stakeholders' power balance. For example, the principle of 'accountability' should be understood as relational and context-dependent, involving priorities and investments of different stakeholders, along with determining who is accountable for decisions and actions to whom, with respect to what. Thus, this study invites a further examination and discussion that would illuminate how these principles should be implemented in a selected context, what information concerns, and what actors they relate to.

Further, the results have shown that there are different actors involved in the context of OPS in higher education; they include students, teachers, institutions, technology companies, regulators, and national authorities. However, even though the actors are identifiable, there is a need for a more comprehensive and relational understanding of the responsibilities and capacities of different actors (Colonna, 2022). For example, previous studies pointed out power imbalance in the OPS (e.g., Balash et al., 2021; Selwyn et al., 2021). In the latter study that examined the use of OPS in Australian universities, Selwyn et al. (2021) ask about the surrender of the information control to commercial technology companies, the hidden labor required to sustain "automated" systems, and the increased vulnerabilities of "remote" studying. That is, this study sets the need to clearly illustrate the role of higher education institutions, and the relationship between these institutions and commercial technology companies. This is one example, but the issue of 'power imbalance' may also concern other types of actors such as students, who may object to using OPS for examination, and educators, who in turn, are guided by the decision to use OPS (in their teaching practice), undertaken by the university.

## 7. Conclusion

The OPS in higher education are evolving rapidly (Han et al., 2022). This paper provided a scoping literature review that examined privacy issues related to the use of OPS in higher education through the lens of the privacy as contextual integrity theory (Nissenbaum, 2010). We argue that the examination of privacy-related issues associated with selected technologies such as OPS should pay considerable attention to the chosen context, in which a studied technology has been used. Further, the application of this theory has been shown to be valuable in investigating privacy in a context that produces a deep and rich understanding of privacy that goes beyond the narrow conceptualization of privacy as control and monitoring mechanisms.

This study contributes to the understanding of privacy issues related to the use of OPS in higher education. The findings provide valuable insights that may be useful for further research of OPS from several perspectives. The discussion of various privacy issues – associated with the adoption of OPS in the higher education context – including types of information, different actors, and transmission principles adds additional value to the development of effective privacy practices in OPS in higher education. The study specifically emphasizes the importance of governing information flows between the key actors in compliance with social norms and privacy principles in a specific context.

# Acknowledgment

The study has been funded by Digital Futures in Stockholm Sweden.